\documentstyle[preprint,aps,version2,epsf]{revtex}  
\def\ttb{t\bar t}
\def\qqb{Q\bar Q}
\def\to{\rightarrow}
\def\RM{{R(0)^2\over 2\pi M}}
\def\MeV{{\rm MeV}}
\def\Lms#1{\Lambda_{\overline{MS}}^{(#1)}}
\def\asp{{\alpha_s\over\pi}}
\def\ams{\alpha_{\overline {MS}} }
\def\amsp{{\ams \over\pi}}
\def\eps{\epsilon}

\newcommand{\oas}{\mbox{${\cal O}(\alpha_s)$ }}
\newcommand{\oasz}{\mbox{${\cal O}(\alpha_s^2)$ }}

  \def\thebibliography#1{{\bf{References}}\list
 {[\arabic{enumi}]}{\settowidth\labelwidth{[#1]}\leftmargin\labelwidth
   \advance\leftmargin\labelsep
   \usecounter{enumi}}
   \def\newblock{\hskip .11em plus .33em minus -.07em}
   \sloppy
   \sfcode`\.=1000\relax}
  
\begin{document}
\def\ttb{t\bar t}
\def\qqb{Q\bar Q}
\def\gg{\gamma\gamma}
\def\to{\rightarrow}
\def\RM{{R(0)^2\over 2\pi M}}
\def\MeV{{\rm MeV}}
\def\Lms#1{\Lambda_{\overline{MS}}^{(#1)}}
\def\asp{{\alpha_s\over\pi}}
\def\ams{\alpha_{\overline {MS}} }
\def\amsp{{\ams \over\pi}}
\def\eps{\epsilon}
\thispagestyle{empty}
\vspace*{-2mm}
\thispagestyle{empty}
\noindent
\hbox to \hsize{
\hskip.5in \raise.1in\hbox{\bf University of Wisconsin - Madison}
\hfill$\vcenter{\hbox{\bf MAD/PH/827}
                \hbox{\bf KA TTP94-2}
                \hbox{\mbox{}}
            }$
               }
\mbox{}
\hfill  March  1994   \\   
\vspace{0.5cm}
\begin{center}
  \begin{Large}
  \begin{bf}
JET PRODUCTION IN
POLARIZED\\
 LEPTON-HADRON SCATTERING\\
  \end{bf}
  \end{Large}
  \vspace{0.8cm}
  \begin{large}
 E. Mirkes$^a$
 and C. Ziegler$^b$\\[5mm]
  \end{large}
{\it
$^a$Physics Department, University of Wisconsin, Madison WI 53706, USA\\[2mm]
$^b$ Institut f\"ur Theoretische Teilchenphysik,
    Universit\"at Karlsruhe,
    Kaiserstr. 12,
    75128 Karlsruhe 1, Germany\\[2cm]

}
  {\bf Abstract}
\end{center}
\begin{quotation}
\noindent
We calculate  exact analytical expressions for \oas  3-jet  and \oasz \\ 4-jet
cross sections in polarized deep inelastic lepton nucleon scattering.
Introducing an invariant jet definition scheme,
we present differential distributions of 3- and 4-jet cross sections
in the basic kinematical variables $x$ and $W^2$ as well
 as total jet cross sections
and show their dependence
on the chosen spin-dependent (polarized) parton distributions.
Noticebly differences in the predictions are found for the two extreme
choices, i.e. a large negative sea-quark density or a large positive gluon
density. Therefore, it may be possible to discriminate between different
parametrizations of polarized parton densities, and hence between
the different physical pictures of the proton spin
underlying these parametrizations.
\end{quotation}
\newpage
\thispagestyle{empty}
\mbox{}
\newpage
\setcounter{page}{1}
\section{Introduction}
The spin structure of the longitudinally polarized proton
has attracted much theoretical interest during the past years.
This is mainly due to the disagreement between the theoretical
prediction of the Ellis-Jaffe sum rule \cite{Ellis-Jaffe}
for the first moment of the polarized structure function
$g_1(x,Q^2)$ and the combined SLAC-EMC-data \cite{SLAC-EMC}.
The experimental results which turn out to be much smaller
than expected suggest two extreme scenarios:
a large negative polarization for
the sea-quarks \cite{Ellis3+1} or alternatively a large positive polarization
of gluons in the longitudinally polarized proton \cite{Altarelli2+2,AltStir}
(see also the discussion in sect.~II).
A combination of both is also possible.

There have been many suggestions for measuring polarized parton
distributions with hard processes \cite{4*Reya} .
Only processes where the polarized gluon distribution
$\bigtriangleup g$ enters at leading order are a promising
source for measuring the polarized parton distributions.

In this paper we study  3- and 4-jet cross sections
in polarized deep inelastic lepton-nucleon processes using recently
proposed polarized quark and gluon distributions, $\triangle q$
and $\triangle g$ respectively.

To \oas one has 3-jet final states\footnote{We include the
target jet when counting the number of produced jets.}.
The relevant subprocesses are
$\gamma^{\ast}q \longrightarrow Gq$ and
$\gamma^{\ast}G \longrightarrow q \bar{q}$.
In \oasz the following subprocesses contribute to the 4-jet production:
$\gamma^{\ast}q \longrightarrow GGq , \gamma^{\ast}q
\longrightarrow qq \bar{q}$ and $\gamma^{\ast}G \longrightarrow q \bar{q}G$.
All partons are assumed to be massless.
We present exact analytical expressions for the relevant
3- and 4-jet partonic hard scattering cross sections,
using polarized beams and polarized targets,
and specify how these partonic cross sections are folded with
the respective polarized parton densities to obtain the DIS
hadronic cross section.
Introducing an invariant jet definition scheme we present
differential distributions of 3- and 4-jet rates.

Large differences in the differential distributions
are found for the two extreme choises for the polarized
parton distributions made in the literature i.e.~a large
negative sea-quark density or a large positive
gluon density.
Thus, the analysis of jet production in polarized deep
inelastic scattering allows to extract $\Delta \bar{q}$
and/or $\Delta G$ from future measurements and it may be possible
to discriminate between the distinctly different physical pictures
of the proton spin underlying these parametrizations.

The paper is organized as follows:
In section II  we introduce our notations and some useful formulae.
In section III we discuss the problem of defining jets in deep inelastic
scattering and present analytical results for the $O(\alpha_s)$ 3-jet cross
sections.
Section IV contains analytical formulae for the $O(\alpha_s^2)$ 4-jet cross
sections.
We specify in detail how the partonic hard scattering cross sections are folded
with the
respective parton distribution functions. In setion V the set of polarized
parton
distribution functions used in our calculations are reviewed.
Section VI contains our numerical results.
We discuss overall kinematical cuts that we choose to define our deep inelastic
scattering phase space. Further cuts are then imposed on the invariant jet-jet
masses
$s_{ij}$ to define the resolved 3- and 4-jet phase space. We present
predictions
for the 3- and 4-jet cross sections and show distributions in the kinematical
variables
$x$ and $W^2$.
Section VII contains our summary. Finally, detailed analytical
formulae for the 4-jet partonic matrix elements are relegated to the appendix.

\section{Kinematics of polarized DIS and the structure function $g_1$}
Consider deep inelastic lepton-proton scattering ($X$ is an
arbitrary hadronic final state)
\begin{equation} \label{disreac}
l_1(l)+H(P) \longrightarrow l_2(l')+X(P_f) \end{equation}
The particle momenta are given within the brackets.

Reaction (\ref{disreac}) proceeds via the exchange of an intermediate vector
boson $V=\gamma^{\ast},Z,W^{\pm}$. In this paper we will only
consider the exchange by a virtual photon.
We denote the
$\gamma^{\ast}$-momentum by $q$, the absolute square by $Q^2$, the
cms energy by s, the square of the final hadronic mass by
$W^2$ and introduce the scaling variables $x$ and $y$:
\begin{eqnarray} \label{defkin} q & = & l-l' \nonumber \\
Q^2 & \equiv & -q^2=xys>0 \nonumber \\
s & = & (P+l)^2 \nonumber \\ W^2 & \equiv & P_f^2=(P+q)^2 \\
x & = & \frac{Q^2}{2Pq} \hspace{1cm} (0<x \le 1) \nonumber \\
y & = & \frac{Pq}{Pl} \hspace{1cm} (0<y \le 1). \nonumber \end{eqnarray}
For fixed s, only two variables in (\ref{defkin}) are independent, since e.~g.
\begin{displaymath} xW^2=(1-x)Q^2,\hspace{1cm}  Q^2=xys. \end{displaymath}
For polarized lepton and polarized target scattering, the
inclusive (with respect to the outgoing hadrons) cross section
for process (\ref{disreac}) can be written as
\begin{equation}  \frac{d \, \Delta \sigma}{dx \, dy} \equiv
\frac{d^2 \sigma^{\uparrow \uparrow}}{dx \, dy}
- \frac{d^2 \sigma^{\uparrow \downarrow}}{dx \, dy}
=\frac{8 \pi \alpha^2}{q^2} \left[ \left\{ 2-y- \frac{Mxy}{E} \right\}
g_1(x,Q^2)- \frac{2Mx}{E}g_2(x,Q^2) \right].
\end{equation}
The left arrow above $\sigma$ denotes the polarization of the
incoming lepton with respect to the direction of its momentum.
The right arrow stands for the polarization of the proton
parallel or anti-parallel to the polarization of the incoming lepton.
$g_1(x,Q^2)$ and $g_2(x,Q^2)$ are the polarized structure functions
for longitudinal and transverse spin respectively.
At large energies, only $g_1$ contributes.\\
In the QCD improved parton model,
the polarized structure function $g_1^P$ of the proton can be
expressed in the lowest-order approach
by\footnote{We follow the notation of Gl\"uck, Reya,
Vogelsang \cite{GRV}.}
\begin{equation} \label{2g1} 2g_1^P(x,Q^2)=\frac{4}{9}\Delta u_v(x,Q^2)
+\frac{1}{9}\Delta d_v(x,Q^2) + \frac{4}{3}\Delta \bar{q}
(x,Q^2)  \end{equation}
We have taken an SU(3)-flavour symmetric polarized
sea  $ \Delta \bar{q} \equiv \Delta \bar{u}
= \Delta\bar{d} = \Delta\bar{s} = \Delta s $
where $ \Delta u_v $ and $\Delta d_v$ are the
polarized valence quark distributions and
\begin{equation} \label{deltaq} \Delta q(x,Q^2)=q_+(x,Q^2)-q_-(x,Q^2).
\end{equation}
In (\ref{deltaq}) $\:q_+  (q_-)$ denotes a parton of flavour $q$
in the proton whose spin is aligned \mbox{(anti-aligned)}
to the proton's spin.\\
Experimental measurements of the asymmetry
\begin{equation}\label{asymmetry}
   A_1^P(x,Q^2) \equiv \frac{2x\,g_1^P(x,Q^2)}{F_2^P(x,Q^2)}
\end{equation}
provide us with a determination of $g_1^P(x,Q^2)$.
In (\ref{asymmetry}) $F_2^P(x,Q^2)$ is the unpolarized structure
function, i.e.~ $F_2^P(x,Q^2) = x\sum\limits_q e_q^2
[q(x,Q^2)+\bar{q}(x,Q^2)]$.\\
Writing $g_1^P(Q^2) \equiv \int\limits_0^1g_1^P(x,Q^2)dx$ as
\begin{equation}
   g_1^P(Q^2) = \frac{1}{2}[\frac{4}{9}\Delta u(Q^2) + \frac{1}{9}
   \Delta d(Q^2) + \frac{1}{9} \Delta s(Q^2)]
\end{equation}
and using the $F/D$ ratio of the hyperon decay and the Bjorken sum rule,
the following values for the first moments of the polarized
quark densities have been obtained \cite{SLAC-EMC,Baum,Altarelli}
\begin {eqnarray*}
   \Delta u & = & 0.78 \pm 0.06\\
   \Delta d & = & -0.47 \pm 0.06\\
   \Delta s & = & -0.19 \pm 0.06
\end{eqnarray*}
and
\begin{equation}
\Delta q \equiv \Delta u + \Delta d + \Delta s = 0.12 \pm 0.17.
\end{equation}

Therefore the total spin carried by the quarks and antiquarks in a polarized
proton is small and actually compatible with zero.
This is the famous EMC result.
 Recently, new data from the SMC muon-deuteron
scattering experiment \cite{SMC} and from the SLAC muon-He scattering
experiment \cite{SLAC}
are available. It is shown in \cite{Ellisneu}, that a global average of the
present existing date yields
 $\Delta q \equiv \Delta u + \Delta d + \Delta s = 0.22 \pm 0.10$
which is still not very different
 from  the value in (\ref{deltaq}).

Several groups have suggested that the difference between the
Ellis-Jaffe sum rule \cite{Ellis-Jaffe} (the sea contribution was set to zero
in the
Ellis Jaffe sum rule) and the EMC result could be resolved by a larger higher
correction due to a large gluon polarization
\cite{Altarelli2+2,AltStir} or by a large sea contribution
\cite{Ellis3+1}. Various sets of helicity
difference distribution functions have been developed based on these ideas.
An independent determination of the gluon spin densities are therefore crucial
in the understanding of the spin structure of the proton.

In the following, we study \oas 3-jet and \oasz 4-jet production in
polarized DIS (using the polarized parton distributions
proposed in \cite{Cheng+Wai}) and show that these processes yield strong
information
on the spin-dependent quark and gluon densities.
\section{Jet cross sections in \oas}
The Quark-Parton model ($ \cal O$ ($ \alpha_s^0 $)) process
\begin{displaymath} l_1(l)+q(p=\eta P) \longrightarrow l_2(l')+q(p_1)
\end{displaymath}
predicts two and only two jets in the final state, namely the
struck quark jet and the hadron remnant jet which leads
to the prediction of two back-to-back jets in the
hadronic center-of-mass-frame.

The \oas tree graph processes (see fig.~1a,b for the relevant boson-parton
subprocesses)
\begin{eqnarray} l_1(l)+\stackrel{(-)}{q}(p) & \longrightarrow &
l_2(l')+\stackrel{(-)}{q}(p_1)+G(p_2) \\
l_1(l)+G(p) & \longrightarrow & l_2(l')+q(p_1)+\bar{q}(p_2)
\end{eqnarray}
give rise to 3-jet production.
There are singular regions in the phase space of the \oas
tree graph cross sections. The singular behaviour corresponds
to the situation where two (or more) partons are irresolvable
and have to be counted as contributions to the 2-jet cross section.
Parts of these singular cross section contributions will cancel
against corresponding singular terms of the \oas virtual 2-parton
corrections.
A remaining initial state mass singularity has to be absorbed
into the polarized parton densities.

In order to calculate a 3-jet cross section we have to {\em define}
what we call 3 jets by introducing a {\em resolution criterion}.
As has been elaborated in detail in \cite{Hera1,Hera2}
energy-angle cuts are
not suitable for an asymmetric machine with its strong
boosts from the hadronic cms to the laboratory frame.
As a jet resolution criterion we use the invariant mass cut
criterion introduced in \cite{Hera1}
such that
\begin{equation} \label{sijycut} s_{ij} \ge M^2_c =
\max \{y_{cut}W^2,M_0^2 \} = \max \{ y_{cut}
(p_1+p_2+p_3)^2,M_0^2 \} \hspace{0.2cm} (i,j=1,2,3;i\not= j )
\label{jetdef}
\end{equation}
where $y_{cut}$ is the resolution parameter and $s_{ij}$
is the invariant mass of any two final state partons,
including the remnant jet with momentum
$ p_3=(1- \eta )P $. 
$M_0$ is a fixed mass cut which we have introduced
in order to clearly separate the perturbative and non-perturbative
regime in the case where $W^2$ is small
(see the detailed discussion in sect. VI).
The 3-jet phase space is represented by the kinematical
variables
\begin{equation} x_p=\frac{Q^2}{2 \, pq}=\frac{x}{\eta}
\qquad z=\frac{pp_1}{pq} \quad\mbox{and}\quad \phi .
\end{equation}
The angle $\phi$ denotes the angle between the parton plane
$(\vec{p},\vec{p}_1)$ and the lepton plane
$(\vec{l}_1,\vec{l}_2)$ in the $\vec{p}+\vec{q}=0$ center of mass frame.

The \oas 3-jet cross section for polarized DIS is then
obtained by
\begin{eqnarray} \label{3-Jet-WQ}
\frac{d^2\Delta\sigma}{dx \, dy}&=&
\frac{2\pi\alpha^2\,(2-y)}{Q^2}\,\frac{\alpha_s(\mu_R^2)}{2\pi}
\int\limits_{\eta_{min}}^{\eta_{max}}\frac{d\eta}{\eta}
\int\limits_{z_{min}}^{z_{max}}dz
\int\limits_{0}^{2 \pi}\frac{d \phi}{2 \pi}\label{jet3hadronic} \\
&&\bigg\{ I_{\Delta q}^{(0)} + I_{\Delta G}^{(0)}
+ \frac{\sqrt{1-y}}{2-y} \cos\phi
\Big( I_{\Delta q}^{(1)}+I_{\Delta G}^{(1)} \Big) \bigg\}.\nonumber
\end{eqnarray}
The integration limits in $\eta$ and $z$ originate
from our jet resolution criterion (\ref{sijycut}) such that
\begin{eqnarray}
z_{min} & = & \frac{M_c^2}{ys(1-\eta)}=1-z_{max}, \\
\eta_{min} & = & x+\frac{M_c^2}{ys},
\hspace{1.5cm} \eta_{max}=1-\frac{2M_c^2}{ys}.
\end{eqnarray}
The 3-jet integrands are given by $(i=0,1)$
\begin{eqnarray} I_{\Delta q}^{(i)} & = &
\sum_{f=1}^{n_f} e_f^2(\Delta q(\eta,\mu_F^2)+\Delta \bar{q}(\eta,\mu_F^2))
\Delta |{\cal M}^{(i)}|^2_{q \to qG}, \\
I_{\Delta G}^{(i)} & = & (\sum_{f=1}^{n_f} e_f^2) \, \Delta G(\eta,\mu_F^2)\;
\Delta |{\cal M}^{(i)}|^2_{G \to q \bar{q}}
\end{eqnarray}
where $\mu_F$ denotes the factorization mass in the polarized
parton densities $\Delta q$ and $\Delta G$.
The 3-jet matrix elements
$\Delta |{\cal M}^{(i)}|^2_{q \to qg}$ and
$\Delta |{\cal M}^{(i)}|^2_{G \to q\bar{q}}$
are given by
\begin{eqnarray} \Delta |{\cal M}^{(0)}|^2_{q \to qG} & = &
C_F[\frac{x_p^2+z^2}{(1-x_p)(1-z)}+2(x_p+z)], \\
\Delta |{\cal M}^{(0)}|^2_{G \to q\bar{q}} & = &
T_R(2x_p-1)(\frac{1}{z}+\frac{1}{1-z}-2),\\
\mbox{\vspace{1em}}\nonumber\\
\Delta |{\cal M}^{(1)}|^2_{q \to qG} & = &
4C_F\sqrt{\frac{x_pz}{(1-x_p)(1-z)}}\:(1-x_p-z),\\
\Delta |{\cal M}^{(1)}|^2_{G \to q\bar{q}} & = &
4T_R\sqrt{\frac{x_p(1-x_p)}{z(1-z)}}\:(1-2z),
\end{eqnarray}
where $C_F=\frac{4}{3}$ and $T_R=\frac{1}{2}$ are the corresponding
colour factors.
Note that the contribution of the matrix elements
$\Delta |{\cal M}^{(1)}|^2$ vanish after integration of the angle
$\phi$ in (\ref{3-Jet-WQ}).
Furthermore, the $z$-integration can be done analytically yielding:
\begin{eqnarray}
\int_{z_{min}}^{z_{max}}
\int_0^{2\pi}\frac{d\phi}{2\pi}
\Delta |M^{(0)}|^2_{q\rightarrow qG} dz &=&
C_F \left\{ -\frac{1+x_p^2}{1-x_p}\ln\frac{z_{min}}{1-z_{min}}\right.\nonumber
\\
&& \left. +\left(2x_p+1-\frac{3}{2}\frac{1}{1-x_p}\right)(1-2z_{min})\right\},
\end{eqnarray}
\begin{eqnarray*}
\int_{z_{min}}^{z_{max}}
\int_0^{2\pi}\frac{d\phi}{2\pi}
\Delta |M^{(0)}|^2_{G\rightarrow q\bar{q}} dz&=&
-T_R (2x_p-1)\left(\ln\frac{z_{min}}{1-z_{min}}+1-2z_{min}\right).
\end{eqnarray*}
\section{Four jet cross sections}
In this section we provide exact \oasz 4-jet production
cross sections in polarized deep
inelastic scattering.
The lowest order contributions to the
$2 \to 4$ parton process $l+p \longrightarrow l'+p_1+p_2+p_3$ are
(the relevant boson-parton subprocesses are depicted in figs.~(2a-c)):
\begin{eqnarray} \label{equ1} l+q & \longrightarrow & l'+q+G+G, \\
   \label{equ2} l+G & \longrightarrow & l'+q+ \bar{q} +G,\\
   \label{equ3} l+q & \longrightarrow & l'+q+q+ \bar{q}
\end{eqnarray}
and antiquark initiated processes.
The corresponding differential partonic cross sections
$d{\Delta\hat{\sigma}}_p$ are given by  ($s_p=(p+l)^2$)
\begin{displaymath} d\Delta\hat{\sigma}_p=\frac{1}{2s_p}
\frac{\Delta\overline{|{\cal M}_i^{4-jet}|^2}}{Q^4} dPS^{(4)}.
\end{displaymath}
Here $\Delta\overline{|{\cal M}_i^{4-jet}|^2}$ denotes
the squared matrix element (without the photonpropagator)
of the parton processes (\ref{equ1})-
(\ref{equ3}) including all coupling, statistical and colour factors.
The lorentz invariant four particle phase space
\begin{displaymath} dPS^{(4)}=(2\pi)^4
\delta^4(l+p-l'-p_1-p_2-p_3) \frac{d^3\vec{l}'}{(2\pi)^3 2E'}
\frac{d^3\vec{p}_1}{(2\pi)^3 2E_1} \frac{d^3\vec{p}_2}{(2\pi)^3 2E_2}
\frac{d^3\vec{p}_3}{(2\pi)^3 2E_3}.
\end{displaymath}
can be parametrized
by seven kinematical variables \cite{Hera2}, which we choose as
$y, Q^2, z', \phi , s_{23}, \cos \theta_{23}$ and $\phi_{23}$
where $y$ and $Q^2$ are already defined in (\ref{defkin}).
$s_{23}$ denotes the squared invariant mass of partons
2 and 3
\begin{displaymath}
s_{23}=(p_2+p_3)^2
\end{displaymath}
and the variable $z'$ is given by
\begin{displaymath}
z'=1- \frac{pp_1}{pq}.
\end{displaymath}
$\phi$ is the azimuthal angle of the outgoing lepton in the
(boson-initial parton) center of mass frame.
Finally $\theta_{23}$ and $\phi_{23}$ denote the polar and
azimuthal angles of parton 2 in the rest frame
$\vec{p}_2 +\vec{p}_3=0$ and $\hat{s}=(p+q)^2$.
In terms of these variables the phase space can
be decomposed as \cite{Hera2}
\begin{equation}
\int\limits_0^1\!d\eta\int \!dPS^{(4)}=
\frac{1}{2^{14}\pi^7} \int\limits_0^1\!dy
\int\limits_0^{ys}\!dQ^2 \int\limits_x^1\!d\eta
\int\limits_0^1\!dz' \int\limits_0^{2\pi}\!d\phi
\int\limits_0^{z'\hat{s}}\!ds_{23}
\int\limits_{-1}^{+1}\!d\cos\theta_{23}
\int\limits_0^{2\pi}\!d\phi_{23}.
\end{equation}
The  \oasz hadronic 4-jet cross section
is then  given by
\begin{equation}
\frac{(2\pi)^2 d^8 \Delta\sigma_H}{dydQ^2d\eta dz'd\phi ds_{23}
d\cos \theta_{23}d\phi_{23}}=
\sum_{i=-n_f}^{n_f} \Big( \frac{\alpha_s(\mu_R^2)}{2\pi} \Big)^2
\frac{\pi \alpha^2}{8 \eta s} \Delta f_i(\eta, \mu_F^2)
\frac{\Delta \overline{|{\cal M}_i^{4-jet}|^2}}{Q^4}
\label{jet4hadronic}
\end{equation}
where the index $i$ refers to the nature of the incoming
polarized parton (gluon: $i=0$, $n_f$ light quarks:
$i=1, \ldots n_f$ and $n_f$ light antiquarks:
$i=-n_f, \ldots -1$).
$\mu_F$ denotes the factorization mass in the polarized
parton densities and $\mu_R$ is the renormalization
mass entering the strong coupling constant.\\
In the next section we explicitly specify the squared
matrix elements $\Delta \overline{|{\cal M}_i^{4-jet}|^2}$
including their respective parton weights $f_i(\eta, \mu_F^2)$
for the partonic processes (\ref{equ1})-(\ref{equ3}).

\subsection{\oasz Four jet matrix elements in polarized $eP$ scattering}
We are now in the position to calculate the squared
matrix elements of the hard scattering process.
In order to exhibit the parton that initiates the
hard scattering process we write down the relevant hard scattering
process with the parton weights included.
We will also explicitly include the colour factors and final state
identical particle factors in our expressions.
In the following we specify the various quark, antiquark
and gluon initiated processes:\\[2mm]

(i) {\bf Quark and antiquark initiated two quark
two gluon processes $(q_f \to q_fGG)$ and
$(\bar{q}_f \to \bar{q}_fGG)$}\\
These processes receive contributions from the
diagrams of fig. 2a.
We divide the contributions into four classes,
$|{\cal M}|^2_{A,B,C_1,C_2}$ distinguished by their
group weight factors (we follow the notation
in \cite{ERT} for $e^+e^-$ jet production):
\begin{tabbing}
\qquad\= A :\quad\= planar abelian graphs: $N_CC_F^2 \:$
 \qquad\qquad\qquad\quad\= 11, 21, 22, 31, 32, 33, \\
\> \> \> 44, 54, 55, 64, 65, 66\\
\> B :\> non-planar abelian graphs: $N_CC_F(C_F-\frac{N_C}{2} \:$)
\> 41, 42, 43, 51, 52, 53\\
\> \> \> 61, 62, 63\\
\> $C_1$:\> QED-QCD interference graphs:
$\frac{1}{2}N_C^2C_F \:$
\> 71, 72, 73, 74, 75, 76,\\
\> \> \> 81, 82, 83, 84, 85, 86\\
\> $C_2$:\> QCD-QCD graphs: $N_C^2C_F \:$
\> 77, 87, 88.
\end{tabbing}
The squared matrix element is then given by:
\begin{equation} \label{matelq} \Delta \overline{|{\cal M}^{4 jet}_
{q_f \to q_fGG}|^2}=\frac{1}{3} \cdot \frac{1}{2}
\left(\sum_{f=1}^{n_f} e_f^2 \left(\Delta q_f+\Delta \bar{q}_f
\right) \right) \otimes \Delta |{\cal M}_{q_f \to q_fGG}|^2_{A+B+C_1+C_2}
\end{equation}
The numerical factors in (\ref{matelq}) are initial state
averaging factor for parton colour and the final state
identical particle factor. The symbol $\otimes$ denotes
the folding with the parton densities.
The numbers $ij$ in the third column refer to the
interference of graph $i$ with $j$  $(\equiv (ij), i \ge j)$ in fig. 2a.
The explicit expressions for the squared matrix elements
${ij}$ can be found in the Appendix.
They are denoted by $d(i,j)$ and include the color, statistical and
average factors.\\[3mm]
(ii) {\bf Gluon initiated process $(G \to Gq_f \bar{q}_f)$}\\
These processes receive contributions from the
diagrams of fig. 2b.
The contributions can be divided in similar classes
as before (Colour factors and associated graphs are
explicitely given):
\begin{tabbing}
\qquad\= A :\quad\= planar abelian graphs: \qquad\qquad
 \qquad\qquad\quad\= 11, 21, 22, 31, 32, 33, \\
\> \>$T_RC_F(N_C^2-1)$ \> 44, 54, 55, 64, 65, 66\\
\> B :\> non-planar abelian graphs:
\> 41, 42, 43, 51, 52, 53\\
\> \> $T_R(C_F-\frac{N_C}{2})(N_C^2-1)$ \> 61, 62, 63\\
\> $C_1$:\> QED-QCD interference graphs:
\> 71, 72, 73, 74, 75, 76,\\
\> \> $\frac{1}{2}T_RN_C(N_C^2-1)$ \> 81, 82, 83, 84, 85, 86\\
\> $C_2$:\> QCD-QCD graphs: $T_RN_C(N_C^2-1)$
\> 77, 87, 88.
\end{tabbing}
The squared matrix element reads
\begin{equation} \Delta \overline{|{\cal M}^{4 jet}_
{G \to Gq_f\bar{q}_f}|^2}= \frac{1}{8} \cdot
\left(\sum_{f'=1}^{n_{f'}} e_{f'}^2 \right) \Delta G
\otimes |{\cal M}_{G \to Gq_f\bar{q}_f}|^2_{A+B+C_1+C_2}
\end{equation}
where $\frac{1}{8}$ is an averaging factor for the initial
gluon colours. The numbers $ij$ in the third column
refer to the interference of graph $i$ with graph $j$
 in fig. 2b.
The explicit expressions for the squared matrix elements
${ij}$
are to long to be presented in this paper. They are available from the
authors.\\[2mm]
(iii) {\bf Quark and antiquark initiated four-quark process
$(q_f \to q_fq_{f'}\bar{q}_{f'})$ and
$(\bar{q}_f \to \bar{q}_fq_{f'}\bar{q}_{f'})$}\\
These diagrams receive contributions from the diagrams
in fig. 2c.
The contributions are divided in four classes
distinguished by their group weight factors:
\begin{tabbing}
\qquad\= D :\quad\= non-interference contributions: $N_CC_FT_R$
 \qquad\= 11, 21, 22, 55, 65, 66\\
\> $\mbox{D}'$ :\> non-interference contributions: $N_CC_FT_R$
\> 33, 43, 44, 77, 87, 88\\
\> E :\> non-interference contributions:
\> 31, 32, 41, 42, 51, 52,\\
\> \> $N_CC_F(C_F-\frac{N_C}{2})$ \> 61, 62, 73, 74, 75, 76,\\
\> \> \> 83, 84, 85, 86\\
\> F :\> interference contributions: $N_CC_FT_R$
\> 53, 54, 63, 64, 71, 72,\\
\> \> \> 81, 82.
\end{tabbing}
The squared matrix element reads
\begin{eqnarray}
\Delta \overline{|{\cal M}^{4 jet}|^2}_{q \to qq\bar{q}} & = &
\frac{1}{3} \cdot \frac{1}{2}
\bigg[ \sum_{f=1}^{n_f} e_f^2 \left( \Delta q_f+\Delta \bar{q}_f
\right) \otimes \left(n_{f'} \cdot |{\cal M}|^2_D
+|{\cal M}|^2_E \right)\nonumber\\
  &  & +\sum_{f=1}^{n_f}\big( \sum_{f'=1}^{n_{f'}} e_{f'}^2\big)
\left(\Delta q_f+\Delta \bar{q}_f \right)
\otimes |{\cal M}|^2_{D'}
\label{eqxx}\\
  &  & +\sum_{f=1}^{n_f} \big(\sum_{f'=1}^{n_{f'}} e_{f'}\big)e_f
\left(\Delta q_f- \Delta \bar{q}_f \right) \otimes
|{\cal M}|^2_F \bigg] \nonumber
\end{eqnarray}
($\frac{1}{3}$ colour averaging, $\frac{1}{2}$ statistical
factor for identical particles in the final state).
The numbers $ij$ in the third column
refer to the interference of graph $i$ with graph $j$
 in fig. 2c.
The explicit expressions for the squared matrix elements
${ij}$ can be found in the Appendix.
They are denoted by $e(i,j)$ and include the color, statistical and
average factors.

Some comments are in order for this subprocess:
We do not distinguish between equal and unequal
flavour cases. The overcounting in the unequal
flavour case $(f\not=f')$ is compensated by
our multiplication of these contribution with the Fermi
statistical factor $\frac{1}{2}$ as in the equal flavour case
$(f=f')$.\\
The interference contributions class $F$ involve the
calculation of two fermion traces.
They involve a product of traces of 3 fermion
(or antifermion) propagators.
Each such trace is anti-symmetric under quark-antiquark
exchange for the vector current $ \gamma_{\mu} $ due
to charge conjugation invariance. Note that the
interference class $F$ does not contribute to the
\underline{total} cross section if one does not
identify the flavours of the produced quarks and antiquarks.
This is a consequence of charge conjugation -- the
non-abelian generalization of Furry's theorem.
\section{Polarized parton densities\label{polpartden}}
The hadronic 3- and 4-jet cross sections are obtained
by folding the partonic cross sections with the
polarized parton densities (see  (\ref{jet3hadronic},\ref{jet4hadronic})).
In this section, we review the set of the
polarized parton density parametrization
\cite{Cheng+Wai} which we will use
in our numerical calculation.
The sea and gluon parton densities are chosen in such a way
that the experimental result for the Ellis-Jaffe sum rule
is reproduced up to \oas.\\
  These densities are given by:
  \begin{eqnarray} \Delta u_V(x) & = & \alpha(x)u_V(x,M^2)\\
  \Delta d_V(x) & = & \beta(x)d_V(x,M^2)
  \end{eqnarray}
  with
  \begin{displaymath} \alpha(x)=x^{0.287} ,
  \hspace{1cm} \beta(x)=\Big(\frac{x-x_0}{1-x_0}\Big)
  z_p \end{displaymath}
  with $x_0=0.75$ and $p=0.76$.
  The $u_V(x,M^2)$ and $d_V(x,M^2)$ are the unpolarized
  parton densities given by the DFLM4 set ($\Lambda$=0.2 GeV)
  at $M^2=10\mbox{GeV}^2$ \cite{DFLM4}.
  According to \cite{Cheng+Wai} we can adopt the
  following parametrization:
  \begin{itemize}
  \item[set 1:] At $M^2=10\mbox{ GeV}^2$ one has a large negative
  polarized sea-quark density and a zero polarized gluon density
  \begin{eqnarray} \Delta s(x) & = & -11.8\:x^{0.94}\nonumber
  (1-x)^5s(x,M^2),\\ \Delta G(x) & = & 0 .\label{g=0}
  \end{eqnarray}
  \item[set 2:] At $M^2=10\mbox{ GeV}^2$ one has a zero polarized
  sea-quark density and a large positive gluon density
  \begin{eqnarray} \Delta s(x) & = & 0 ,\nonumber\\
  \Delta G(x) & = & 6.0\:x^{0.76}(1-x)^3G(x,M^2) \label{g=max}.
  \end{eqnarray}
  The above gluon density is about the same as the one proposed
  in \cite{AltStir}.
  \item[set 3:] In a realistic situation it is unlikely
  that $\Delta s(x)$ or $\Delta G(x)$ vanishes at some
  scale for all $x$. Therefore the authors in \cite{Cheng+Wai}
  also proposed
  \begin{eqnarray} \Delta s(x) & = & -3.39\:x^{0.62}(1-x)^{1.4}
  s(x,M^2) ,\nonumber\\
  \Delta G(x) & = & 2.69\:x^{0.76}(1-x)^3G(x,M^2) ,\label{g=mid}
  \end{eqnarray}
  for $M^2=10\,\mbox{GeV}^2$.
  \end{itemize}
  $s(x,M^2)$ and $G(x,M^2)$ represent the unpolarized
  parton densities belonging to the DFLM4 set \cite{DFLM4}.

\section {Numerical results}
We will now turn to our numerical cross section results.
In order to define the deep inelastic scattering region,
we choose the lower bounds
\begin{eqnarray} \label{QWx>=} Q^2 & \ge & Q_0^2=4\mbox{ GeV}^2,\nonumber\\
W^2 & \ge & W_0^2=5\mbox{ GeV}^2,\\
x & \ge & x_0\equiv 10^{-2}.\nonumber
\end{eqnarray}
We introduce a limit in $W^2$ to insure an appropriate
hadronic final state.
When producing well separated multijet final states
the effective lower limit of $W^2$ will actually
be increased (see  (\ref{Weta3},\ref{Weta4})).
As already discussed in section III one encounters
infrared/mass (IR/M) singularities when calculating
the \oas and \oasz tree graph cross sections, which
originates from the emission of soft and/or collinear partons.
The tree graph cross sections make sense only for the
production of well separated hard jets.
We introduce the following invariant mass-cut-off
to define the finite tree graph cross sections\footnote{
Note, that the remnant jet
($p_3$ in $\oas$ and $p_4$ in $\oasz$) is included in this definition.}
 (see  (\ref{jetdef})):
\begin{equation} \label{sijpij} s_{ij}\ge
\max\{y_cW^2, M_0^2\}, \hspace{1cm} s_{ij}\equiv
(p_i+p_j)^2 \hspace{0.5cm} i\neq j=1,2,3,(4)
\end{equation}
where $y_c$ is the resolution parameter for which we choose
$y_c=0.04\,\,$ \cite{Hera1,Hera2}.
$M_0$ is an additional fixed mass cut which we have introduced in
order to clearly separate the perturbative and non-perturbative
regime in the case, where $W^2$ is small. Note that
$W^2$ is limited from below by $W_0^2$ (see (\ref{QWx>=})).
This leads to too small invariant masses for a perturbative
calculation to be meaningful. For example with $y_c=0.04$
one obtains $s_{ij}\le0.2 \mbox{ GeV}^2$.
A reasonable choice for the fixed mass cut in (\ref{sijpij})
is $M_0=2\mbox{ GeV}$ \cite{Hera2}.
Our subsequent analysis is then based on the invariant mass cut
\begin{equation}\label{sij**} s_{ij}\ge\max\{0.04\:W^2, 4\mbox{ GeV}^2\}.
\end{equation}
The invariant mass cut in (\ref{sij**}) introduces then the
following available ranges in $\eta$ and $W^2$ for the \oas
3-jet and \oasz 4-jet cross sections:\\
\underline{\oas 3-jet case}:
\begin{eqnarray}\label{Weta3} W^2 & \ge \max\{W_0^2, 3\:M_0^2\}
& =12\:\mbox{GeV}^2 \nonumber\\
\eta & \ge y_c+x(1-y_c) &
\end{eqnarray}
\underline{\oasz 4-jet case}:
\begin{eqnarray} \label{Weta4} W^2 & \ge \max\{W_0^2, 6\:M_0^2\}
& =24\:\mbox{GeV}^2 \nonumber\\
\eta & \ge 3\: y_c+x(1-3\:y_c) &
\end{eqnarray}
As can be seen from (\ref{Weta3}) and (\ref{Weta4}), the
lower bound in $\eta$ and $W^2$ increases with jet multiplicity.

In our numerical calculation we
 use the one-loop formula of the strong coupling constant
\begin{displaymath}
   \alpha_s(\mu_R^2) = \frac{12\pi}{(33-2n_f)\,\ln\frac{\mu_R^2}{\Lambda^2}}
\end{displaymath}
and the $\Lambda$ value consistent to our choise of polarized
parton distributions.
The renormalization scale $\mu_R^2$ and the factorization scale $\mu_F^2$
are both fixed to $Q^2$.
The value of $\alpha_s$ is matched at the thresholds $q=m_q$ and the
number of flavours $n_f$ in $\alpha_s$ is fixed by the number
of flavours for which the masses are less than $Q^2$.
Furthermore the number of quark flavours that can be pair-produced
are set equal to $n_f$ chosen in $\alpha_s$.

In figs. 3 a-f  we show how the various cross sections vary with
a given invariant energy $\sqrt{s}$
 for the three different sets of
 parametrizations (set 1-3)
 of the polarized parton densities. The $Q^2$ range is
chosen as $5\mbox{ GeV}^2\leq Q^2 \leq s$.
One observes, that all quark initiated subprocesses are positiv
and have a maximum around $\sqrt{s}=15-20\,\, \mbox{GeV}$.
In contrast, the gluon initiated
contribution to $\Delta\sigma$
 is negativ and decreasing with increasing $\sqrt{s}$.

In figs.~4-7 we show the dependence of the $\oas$ 3-jet cross sections
on the basic kinematical variables $x$ and $W^2$ for the
different parametrizations.
We have restricted the $Q^2$ range from $5\mbox{ GeV}^2
 < Q^2 < 7\mbox{ GeV}^2$ (figs. 4,6) and $9\mbox{ GeV}^2
 < Q^2 < 11\mbox{ GeV}^2$ (figs. 5,7).
The quark (dashed) and gluon (dotted)
 initiated cross sections are shown separately as well
as the sum (solid).

Comparing  figs.~4a-c (5a-c) one
 observes large differences in the $x$ distribution
using different parametrizations.
In particular, figs. 4b (5b) show that
the 3-jet cross section (solid curve) is negative for a large
gluon contribution for $x$ values less than about $ 0.04$.
An analysis of the $x$-dependence at lower $x$ values is therefore
very sensitive to the polarized gluon densities.

Figs. 6a-c (7a-c) show the $W$-distribution of 3-jet production
 for the three different sets of
parametrizations, where the $Q^2$ range is given by
 $5\mbox{ GeV}^2
 < Q^2 < 7\mbox{ GeV}^2$ ($9\mbox{ GeV}^2
 < Q^2 < 11\mbox{ GeV}^2$).
  Here larger values of $W$ are more sensitive to the
 choice of polarized parton distributions. The gluon process gives a
 negative contribution  and dominates the resulting
cross section in the large $W$ range in figs 6b,c.
 This is a reflection of the $x$
behaviour since $W^2=(1-x)ys$.
Like for the $x$-distribution,
 the predictions for the two extreme choices of parton
distributions  are very different.

Figs.~8 and 9
 show  predictions of the $x$ and $W$ dependence
of the $O(\alpha_s^2)$ 4-jet cross sections
 for the three different sets of
parametrizations and
 $5\mbox{ GeV}^2
 < Q^2 < 7\mbox{ GeV}^2$.
The behaviour of the $x$ and $W$ dependence is similar to the $O(\alpha_s)$
 3-jet
cross section. One observes that the $q\rightarrow qGG$ subprocess
dominates the 4 quark process $q\rightarrow qq\bar{q}$ over the whole $x$ and
$W$
 ranges.
 However, for set 2 of the
parton distributions (fig. 8b), the gluon process $G\rightarrow q\bar{q}G$
gives a large negative
contribution for smaller $x$ values, which gives again
 a negativ cross section for the sum of all contributions.
 The sum of the contributions for set 3 of the parton distributions
(figs. 8c, 9c) remains
still positive indicating that the 4-jet cross sections are less sensitive
to the different parametrizations.
The behaviour of the $W$ dependence in the 4-jet distributions
is again similar to the  behaviour found for the 3-jet cross setcions.
The gluon process gives a negative contribution at larger $W$ values.
\section{Conclusions}
Jet production in DIS scattering experiments, using polarized
beams and polarized targets, has been studied.
It is shown that the jet cross sections are strongly
dependent on the particular
form of the spin dependent parton distributions used in the calculation.
Therefore,
 jet production in polarized DIS can be used to put
rigid constraints on the various parametrizations  of polarized
structure functions.
These informations allow to discriminate between the different
physical pictures of the proton spin underlying these parametrizations.\\[5mm]
{\bf Acknowledgements}\\
This work is supported in part by the U.S. Department of Energy under
contract No. DE-AC02-76ER00881,
in part by BMFT Contract 055KA94P1,
 and in part by the University of Wisconsin
Research Committee with funds granted by the Wisconsin Alumni Research
Foundation.
\newpage

\appendix{Matrix elements}
We give here explicit formulae for the squared matrix elements
${\Delta \overline{|{\cal M}_i^{4-jet}|^2}}$ in  (\ref{jet4hadronic}-
\ref{eqxx}).
As usual,
${\Delta \overline{|{\cal M}_i^{4-jet}|^2}}$ can be written in terms
of the lepton tensor $L_{\mu\nu}$ and the hadron tensor
$\Delta H^{\mu\nu}=\frac{1}{2}(H^{\mu\nu\,\,+}-H^{\mu\nu\,\,-})$, where
$H^{\mu\nu\,\,+} (H^{\mu\nu\,\,-}) $ denotes the hadron tensor
for the process when the incoming parton has positiv (negativ) helicity.
The hadron tensor
 $\Delta H^{\mu\nu}$ can be expanded in terms of the following tensors:
\begin{equation}
   \begin{array}{rcl}
   [p,p_1] & := & \varepsilon^{\mu\nu\alpha\beta}p_{\alpha}p_{1\,\beta}
                  \\{}
   [p,p_2] & := & \varepsilon^{\mu\nu\alpha\beta}p_{\alpha}p_{2\,\beta}\\{}
   [p,p_3] & := & \varepsilon^{\mu\nu\alpha\beta}p_{\alpha}p_{3\,\beta}\\{}
   [p_1,p_2] & := & \varepsilon^{\mu\nu\alpha\beta}p_{1\,\alpha}
   p_{2\,\beta}\\{}
   [p_1,p_3] & := & \varepsilon^{\mu\nu\alpha\beta}p_{1\,\alpha}
   p_{3\,\beta}\\{}
   [p_2,p_3] & := & \varepsilon^{\mu\nu\alpha\beta}p_{2\,\alpha}p_{3\,\beta}{}.
   \end{array}
\end{equation}
Furthermore, we introduce the following notation
\begin{equation} \label{stuxyzDef}
   \begin{array}{ccl}
   s & := & pp_3\\
   t & := & p_2p_3\\
   u & := & pp_2\\
   x & := & p_1p_3\\
   y & := & pp_1\\
   z & := & p_1p_2 .
   \end{array}
\end{equation}
To obtain the full matrix elements
${\Delta \overline{|{\cal M}_i^{4-jet}|^2}}$, the corresponding hadron tensors
for each subprocess have to be contracted with
 the anti-symmetric part of the
the lepton tensor $L_{\mu\nu}^{A}$:
\begin{equation}
L_{\mu\nu}^{A} = 2i\epsilon_{\mu\nu\alpha\beta}\,l^{\alpha}l'^{\beta}
\end{equation}
\subsection{ Matrix elements for
 $(q_f \to q_fGG)$  and $(\bar{q}_f \to \bar{q}_fGG)$}
The results for the interference of graph
 $i$ with graph $j$ in fig. 2a is denoted by $d(i,j)$. Note that  $d(i,j)$
contain
the  color, statistical and
average factors as  given in  (\ref{matelq}).
\begin{eqnarray*}
d(1,1)  & = &   - 16 i  [p_2,p_3]   / (9 s z)  \\
d(2,1)  & = &  16 i \Big( [p,p_1] u
   +
    [p,p_2]  (s - x + y + 2 u - t) +  [p,p_3]  u\\
   &  & +  [p_1,p_3]  u
   +
    [p_2,p_3]   u\Big) / (9 s z (x + z + t))  \\
d(2,2)  & = &   - 16 i \Big( -  [p,p_1] (z + t) +  [p,p_2]  x -
    [p,p_3]  (z + t)\Big) / (9 z (x+z+t)^2)  \\
d(3,1)  & = &  16 i \Big( [p,p_1] x +
    [p,p_2]  x +  [p_1,p_2] x +  [p_1,p_3]  (2 x - y + z - u + t)\\
    & & +[p_2,p_3]   x\Big) / (9 s z (s + u - t))  \\
d(3,2)  & = &  16 i \Big( [p,p_1] (s y + s
   u + x^2 - 3 x y + x z - 2 x u + 2 x t + 2 y^2 - y z + 3 y u\\
   & & - 2 y t + u^2 - 2 u t + t^2)
   +  [p,p_2]  ( - 2 x y - x u + y^2
   - y z + y u)\\
   & &  +  [p,p_3]  ( - x y + y^2 + y u - y t + z u - z t)
   +  [p_1,p_2] (s x + s t - x y\\
   & &  + y^2 + y u - y t) +  [p_1,p_3]
   (s y - x y - x u + y^2 + 2 y u)\\
   &  & +  [p_2,p_3]   (s y + s z - x y - x
   u - y z + y u)\Big) / (9 s z (s+u-t) (x+z+t))  \\
d(3,3)  & = &  16 i \Big( [p,p_1] (s - t) +  [p_1,p_2] (s - t) -
    [p_1,p_3]  u\Big) / (9 s (s+u-t)^2)  \\
d(4,1)  & = &   - 2 i \Big( [p,p_1] y (s - x + 2 y - z + u)
   +  [p,p_2]  y (s - x + y)\\
   & &  +  [p,p_3]  y (y - z + u)  +  [p_1,p_2] y (s - x + y)\\
   & &   +  [p_1,p_3]  y (y - z + u)\Big) / (9 s x z u)  \\
d(4,2)  & = &  2 i \Big( [p,p_1] (s z + z u + u t)
   +  [p,p_2]  (s z + x z - x u + y z)\\
   &  & +  [p,p_3]  (s z - 2 x y + x z - x u - y z - y t) +  [p_1,p_2] ( - x u
   + y t)\\
   &  &  -  [p_1,p_3]  x u +  [p_2,p_3]   y (x + z)\Big) / (9 x z u (x + z +t))
 \\
d(4,3)  & = &  2 i \Big( [p,p_1] (s x + s z - x t) -  [p,p_2]  x u +
    [p,p_3]  ( - x u + y t)\\
   & &  +  [p_1,p_2] (s y + s z + s u - x u + 2 y u - y t)
   +  [p_1,p_3]  ( - s y + s z\\
   & &  + s u - x u) -  [p_2,p_3]   y (s + u)\Big) /
   (9 s x u (s + u - t))  \\
d(5,2)  & = &  4 i \Big( - [p,p_1] t +  [p,p_2]  (x + z)
   +  [p,p_3]  (x +z)\Big) /  (9 x z (x + z + t))  \\
d(5,3)  & = &   -2 i  \Big( [p,p_1] (s^2 - 2 s x +3 s y - s z + s u
   - 2 s t + x^2 - 3 x y + x z\\
   & &  - x u + 2 x t + 2 y^2 - y z + y u - 2 y t) +  [p,p_2]  y (s - x + y)\\
   & &  +  [p,p_3]  y (s - x + y) +
   [p_1,p_2] y (s - x + y)\\
   & &  +  [p_1,p_3]  y (s - x + y)\Big) / (9 s x (x+z+t) (s+u-t))  \\
d(6,3)  & = &  4 i \Big( [p,p_1] t +  [p_1,p_2] (s + u)
   +  [p_1,p_3]  (s + u)\Big) /  (9 s u (s + u - t))  \\
d(7,1)  & = &  i \Big( [p,p_1] (2 s x + s z + s t - x^2 + 2 x y - 2 x z
   + x u - x t + 4 y z + 2 y t\\
   & &  + 2 z u - z t - u t) +  [p,p_2]  (s x + 2 s z - x^2 + x y - x z
   + 2 x u - x t\\
   & & + 2 y z + y t + 4 z u - 2 z t) +  [p,p_3]  ( - s z + 2 x y + 2 x z\\
   & &  + 2 x u + y z + 2
   z u) + 2  [p_1,p_2] s (x + t)\\
   & &  +  [p_1,p_3]  ( - 2 s z + x y - x u + 2 y z
   + 3 y t + z u - u t)\\
   & &  +  [p_2,p_3]   (y z + 2 y t + 2 z u)\Big) / (s z t
   (x + z + t))  \\
d(7,2)  & = &  2 i \Big( [p,p_1] ( - x z - x t + z^2 - 2 z t- t^2)
   +  [p,p_2]  (x^2 + x t + z^2 - z t)\\
   & &  + 2  [p,p_3]  (x^2
   + 2 x z + x t + 2 z^2 + z t)\Big) / (z t (x+z+t)^2)  \\
d(7,3)  & = &   -i  \Big( [p,p_1] (s^2 - s x + 2 s y + 2 s z + 3 x u
   - 2 y u - 2 y t + 2 z u - z t\\
   & &  - u^2 + 4 u t - 2 t^2) +  [p,p_2]  (s y + 2 x u - y u - 3 y t)\\
   & &  +  [p,p_3]  (s y - y
   u+ y t - 2 z u + 2 z t)  +  [p_1,p_2] ( - 3 s x - s z - 3 s t\\
   & &  - 2 y
    u + 2 y t) +  [p_1,p_3]  ( - s y + 3 x u - 3 y u - y t + z u + u t)\\
   & &   +  [p_2,p_3]   ( - s y - 2 s z + 2 x u - y u - 3 y t)\Big) / (s t
   (s+u-t) (x+z+t))  \\
d(7,7)  & = &   -i  \Big( [p,p_1] (2 x^2 + 14 x z + 9 x t
   + 4 z^2 + 11 z t - t^2)  +  [p,p_2]  ( - 5 x^2 + 3 x z\\
   & &  + 3 x t + 8 z t)
   +  [p,p_3]  (5 x z + 8 x t - 3 z^2 + 5 z t)\Big)
    / (t^2 (x+z+t)^2)  \\
d(8,1)  & = &  -i\Big([p,p_1]
  (2 s x - 4 s y + s z - 2 s u + s t + x u + x t - 2 y u +
  2 y t + 2 z u\\
  & & - z t - u^2 + u t) + [p,p_2]  (s x - 2 s y - 2 s
  z - x u + x t - y u + 3 y t)\\
  & & + 2 [p,p_3]  z (u - t) + [p_1,p_2]  (
  2 s x - s y - s z + 2 s u + 2 x u - 2 y u)\\
  & &  + [p_1,p_3]  (4 s x - 2 s
  y + 2 s z - s u + 2 s t + 2 x u - y u + y t + z u - u^2 + u  t)\\
  & &  + [p_2,p_3]  (2 s x - s y + 2 y t)\Big)/(s z t (s + u - t))\\
d(8,2)  & = & i\Big([p,p_1]  (2 s x + 2 s z + s t - x^2 + 2 x y + 3 x u - 4 x t
  - 2 y z - 2 y t + z^2\\
  & &  - z u - 2 t^2) + [p,p_2]  (s x + 3 x y +
  3 x u - x t + y z - y t) + [p,p_3]  ( - s z\\
  & & + 2 x y + 2 y t - 3 z
  u + 3 z t) + [p_1,p_2]  ( - 2 s x - 2 s t + x y - y z + y t)\\
  & &  + [p_1,p_3] (x y + 2 x u - y z - 3 y t)
  + [p_2,p_3]  ( - 2 s z + x y + 2 x
  u + y z - 3 y t)\Big)\\
  & & \cdot 1/(z t (s+u-t)(x+z+t))\\
d(8,3)  & = &  -2i\Big([p,p_1]  (s^2 - s u + 2 s
  t + u t - t^2) + 2 [p_1,p_2]  (2 s^2 + 2 s u - s t + u^2 - u
  t )\\
  & &  + [p_1,p_3]  (s^2 + s t + u^2 - u t)\Big)/(s t (s+u-t)^2)\\
d(8,7)  & = &  - i\Big([p,p_1]  (4 s x +
  14 s z + s t + 14 x u - x t - 16 y t + z t - u t + 2 t^2)\\
  & &  + [p,p_2] ( - 5 s x + 7 x u + 5 x t - 8 y t) + [p,p_3]  (5 s z - 8 y
  t - 7 z u + 7 z t)\\
  & &  + [p_1,p_2]  ( - 5 s x + 7 s z - 5 s t - 8 y t)
  + [p_1,p_3]  (5 x u - 8 y t - 7 z u - 7 u t)\\
  & &  + 12 [p_2,p_3]  ( - s z +
  x u)\Big)/(t^2 (s+u-t)(x+z+t))\\
d(8,8)  & = &  - i\Big([p,p_1]  (2 s^2 + 14 s u - 9 s t + 4
  u^2 - 11 u t - t^2)\\
  & &  + [p_1,p_2]  ( - 5 s^2 + 3 s u - 3 s t - 8 u t)
  + [p_1,p_3]  (5 s u - 8 s t - 3 u^2 - 5 u t)\Big)\\
  & & \cdot 1/(t^2 (s+u-t)^2).
\end{eqnarray*}
The remaining matrix elements can be obtained by the
interchange of\\
 $p_2\leftrightarrow p_3, s\leftrightarrow u, x\leftrightarrow z$
from the above results.
\begin{equation}
   \begin{array}{lcrclcrclcl}
   d(4,4) & \longleftrightarrow & d(1,1) & &  d(5,4)
   & \longleftrightarrow &
   d(2,1) & &  d(5,5) & \longleftrightarrow & d(2,2)\\
   d(6,4) & \longleftrightarrow & d(3,1) & &  d(6,5)
   & \longleftrightarrow &
   d(3,2) & &  d(6,6) & \longleftrightarrow & d(3,3)\\
   d(5,1) & \longleftrightarrow & d(4,2) & &  d(6,1)
   & \longleftrightarrow &
   d(4,3) & &  d(6,2) & \longleftrightarrow & d(5,3)\\
   d(7,4) & \longleftrightarrow & d(7,1) & &  d(7,5)
   & \longleftrightarrow &
   d(7,2) & &  d(7,6) & \longleftrightarrow & d(7,3)\\
   d(8,4) & \longleftrightarrow & d(8,1) & &  d(8,5)
   & \longleftrightarrow &
   d(8,2) & &  d(8,6) & \longleftrightarrow & d(8,3).
   \end{array}
\end{equation}

\subsection{\bf Matrix elements for
  $(G \to Gq_f \bar{q}_f)$}
The result for the interference of graph
 $i$ with graph $j$ in fig. 2b is denoted by $c(i,j)$.
The analytical results are to long to present here.
They are available from the authors.

\subsection{ Matrix elements for
$(q_f \to q_fq_{f'}\bar{q}_{f'})$ and
$(\bar{q}_f \to \bar{q}_fq_{f'}\bar{q}_{f'})$}
The results for the interference of graph
 $i$ with graph $j$ in fig. 2c is denoted by $e(i,j)$.
 Note that  $e(i,j)$ contain
the  color, statistical and
average factors as given  in  (\ref{eqxx}).
\begin{eqnarray*}
e(1,1) & = &  - 2 i \Big([p,p_1]  ( - 2 x z - x t - z t) + [p,p_2]  (x
^2 - x z - z t)\\
& &  + [p,p_3]  ( - x z - x t + z^2)\Big)/(3 t^2 (x+z+t)^2) \\
e(2,1) & = & 2 i \Big(2 [p,p_1]
(
 - s z - x u + y t) + [p,p_2]  (s x - x t - x u + y t) \\
& & + [p,p_3]( - s z + y t - z t + z u) + [p_1,p_2]  (s x - s z + s t +
 y t)\\
& &  + [p_1,p_3]  ( - x u + y t + z u + t u)\\
& &  + 2 [p_2,p_3]  (s z -
 x u)\Big)/(3 t^2 (x+z+t) (s-t+u)) \\
e(2,2) & = &  - 2 i \Big([p,p_1]  (s t - 2 s u + t u) +
[p_1,p_2]
  (s^2 - s u + t u) \\
& & + [p_1,p_3]  (s t - s u + u^2)\Big)/(3 t^2 (s-t+u)^2) \\
e(3,1)
 & = & 2 i \Big([p,p_1]  z (x + t) + [p,p_2]  z (x + t) - [p_1,p_3]  z (y +u)\\
& &  - [p_2,p_3]  z (y + u)\Big)/(9 t u (x+z+t) (y-z+u)) \\
e(3,2) & = & 2 i \Big([p,p_1]  ( - x u
+
 y t) + [p,p_2]  ( - s z - x u + y t) + [p,p_3]  z ( - t + u)\\
& &  + [p_1,p_2]  s (y + u) + [p_1,p_3]  (s z - x u + y t)\\
& &  + [p_2,p_3]  (
 - x u + y t)\Big)/(9 t u (s-t+u) (y-z+u)) \\
e(3,3) & = & 2 i \Big([p,p_3]  ( - y z + y u + z^2)
 - [p_1,p_3]  u (y + z)\\
& &  + [p_2,p_3]  (y^2 - y z - z u)\Big)/(3 u^2
(y-z+u)^2) \\
e(4,1) & = & 2 i \Big(
[p,p_1]
  ( - s z - x u + y t) + [p,p_2]  ( - x u + y t) + [p,p_3]
  z ( - t + u)\\
&  &  + [p_1,p_2]  s (x + t) + [p_1,p_3]  ( - x u + y t)\\
& &  + [p_2,p_3]  (s z - x u + y t)\Big)/(9 t u (x+z+t) (s-t+u))\\
e(4,2) & = &  - 4 i \Big([p,p_1]  s
(
t - u) + [p_1,p_2]  s^2 + [p_1,p_3]  s (t - u)\Big)/(9 t u
(s-t+u)^2) \\
e(4,3) & = & 2 i \Big([p,p_1]
 ( - x u + y t - z t + t u) - 2 [p,p_2]  x u\\
& &  + [p,p_3]  (s z -
x u - z t + z u) + [p_1,p_2]  (s y - s z + s u + x u)\\
& &  + [p_2,p_3]
 ( - s y - x u + y t - y u)\Big)/(3 u^2 (s-t+u) (y-z+u))\\
e(4,4) & = &  -2 i \Big([p,p_1]  (s
t
 - s u - t^2) + [p_1,p_2]  (s^2 - s t - t u)\\
& &  - [p_1,p_3]  u
 (s + t)\Big)/(3 u^2 (s-t+u)^2)\\
e(5,1) & = &  -4 i \Big([p,p_1]  z (x + t) + [p,p_2]  z (x + t) -
[p,p_3]  z^2\Big)/(9 x t (x+z+t)^2)\\
e(5,2) & = &  - 2 i \Big([p,p_1]  ( - s z - x u + y t) + [p,p_2]  (
-
 x u + y t) + [p,p_3]  z ( - t + u)\\
& &  + [p_1,p_2]  s (x + t) +
[p_1,p_3]
  ( - x u + y t)\\
& &  + [p_2,p_3]  (s z - x u + y t)\Big)/(9 x t
(x+z+t) (s-t+u))\\
e(5,3) & = & 2 i \Big([p,p_1]  ( - x z - 2 z t + t u) + [p,p_2]  (s z - x z -
x u - z t)\\
& &  + [p,p_3]  (s z - x u - y t) + [p_1,p_2]  ( - s z + x
u - y t)\\
& &  + [p_1,p_3]  ( - s z + x u + y z + z u)\\
& &  + [p_2,p_3]  ( - x
 y + z u)\Big)/(3 x u (x+z+t) (y-z+u))\\
e(5,4) & = &  -2 i \Big([p,p_1]  ( - s z - x u + y t) +
[p,p_2] ( - x t - x u + y t - z t)\\
& &  + [p,p_3]  ( - x t - 2 z t +
 z u) + [p_1,p_2]  (s x - t u)\\
& &  + [p_1,p_3]  ( - s t - x u\\
& &  + y t-t u) + [p_2,p_3]  (s z - x u + y t)\Big)/(3 x u (x+z+t) (s-t+u))\\
e(6,2
)
 & = &  -2 i \Big([p,p_1]  s (t - u) + [p,p_2]  s (x - y) + [p_1,p_3]  s (
t - u)\\
& &  + [p_2,p_3]  s (x - y)\Big)/(9 x t
(s-t+u) (s-x+y))\\
e(6,3) & = &  -2 i \Big([p,p_1]  ( - s z
 + x u - y t) + [p,p_2]  (s y - x y + x u - y t)\\
& &  + [p,p_3]  ( -
 x z + y u) + [p_1,p_2]  ( - 2 s y - s u + x y)\\
& &  + [p_1,p_3]  (x u + y
 z - y t - y u)\\
& &  + [p_2,p_3]  (s z + x u - y t)\Big)/(3 x u
(y-z+u) (s-x+y))\\
e(6,4) & = &  - 2 i \Big(
[p,p_1]  ( - s u + x t) + [p,p_2]  (s x - s y + s z - x u)\\
& &  + [p,p_3]  (s z - x u + y t) + [p_1,p_2]  ( - s z + x u + y t)\\
& &  + [p_1,p_3]
  ( - s z + s t - s u + x u)\\
& &  + [p_2,p_3]  (s x - 2 s y - y u
)
\Big)/(3 x u (s-t+u) (s-x+y))\\
e(7,3) & = & 4 i \Big( - [p,p_3]
  z^2 + [p_1,p_3]  z (y + u) + [p_2,p_3]  z (y + u)\Big)/(9 y u
(y-z+u)^2) \\
e(7,4) & = &  -2 i \Big([p,p_1]
  ( - x u + y t) + [p,p_2]  ( - s z - x u + y t)\\
& &  + [p,p_3]  z
( - t + u) + [p_1,p_2]  s (y + u) + [p_1,p_3]  (s z - x u + y t)\\
& &  + [p_2,p_3]  ( - x u + y t)\Big)/(9 y u
(s-t+u) (y-z+u))\\
e(8,4) & = &  -2 i \Big([p,p_1]  s (t - u) + [p,p_2]
s
 (x - y) + [p_1,p_3]  s (t - u)\\
& &  + [p_2,p_3]  s (x - y)\Big)/(9 y u
(s-t+u) (s-x+y)).\\
\end{eqnarray*}
The remaining matrix elements can be obtained by the
interchange of\\
 $p_1\leftrightarrow p_2, t\leftrightarrow x, u \leftrightarrow y$
from the above results.
\begin{equation}
   \begin{array}{lclclcrclcr}
   e(5,5) & \longleftrightarrow & e(1,1) & & e(6,5) & \longleftrightarrow &
   e(2,1) & & e(6,6) & \longleftrightarrow & e(2,2)\\
   e(7,5) & \longleftrightarrow & e(3,1) & & e(7,6) & \longleftrightarrow &
   e(3,2) & & e(7,7) & \longleftrightarrow & e(3,3)\\
   e(8,5) & \longleftrightarrow & e(4,1) & & e(8,6) & \longleftrightarrow &
   e(4,2) & & e(8,7) & \longleftrightarrow & e(4,3)\\
   e(8,8) & \longleftrightarrow & e(4,4) & & e(6,1) & \longleftrightarrow &
   e(5,2) & & e(7,1) & \longleftrightarrow & e(5,3)\\
   e(8,1) & \longleftrightarrow & e(5,4) & & e(7,2) & \longleftrightarrow &
   e(6,3) & & e(8,2) & \longleftrightarrow & e(6,4)\\
   e(8,3) & \longleftrightarrow & e(7,4).\\
   \end{array}
\end{equation}


\def\npb#1#2#3{{\it Nucl. Phys. }{\bf B #1} (#2) #3}
\def\plb#1#2#3{{\it Phys. Lett. }{\bf B #1} (#2) #3}
\def\prd#1#2#3{{\it Phys. Rev. }{\bf D #1} (#2) #3}
\def\prl#1#2#3{{\it Phys. Rev. Lett. }{\bf #1} (#2) #3}
\def\prc#1#2#3{{\it Phys. Reports }{\bf C #1} (#2) #3}
\def\pr#1#2#3{{\it Phys. Reports }{\bf #1} (#2) #3}
\def\zpc#1#2#3{{\it Z. Phys. }{\bf C #1} (#2) #3}
\def\ptp#1#2#3{{\it Prog.~Theor.~Phys.~}{\bf #1} (#2) #3}
\def\nca#1#2#3{{\it Nouvo~Cim.~}{\bf A #1} (#2) #3}

\newpage
\noindent
{\bf{Figure captions}}
\begin{itemize}
\item[{\bf Fig. 1}]
Diagrams for $\gamma^{\ast}q\rightarrow Gq$ (a) and
             $\gamma^{\ast}G\rightarrow q\bar{q}$ (b).
\item[{\bf Fig. 2}]
Diagrams for $\gamma^{\ast}q\rightarrow qGG$ (a),
             $\gamma^{\ast}G\rightarrow q\bar{q}G$ (b) and
             $\gamma^{\ast}q\rightarrow q\bar{q}q$ (c).
\item[{\bf Fig. 3}]
Dependence of the various $O(\alpha_s)$ (a-c) and $O(\alpha_s^2)$ (e-f)
processes on the total cms energy $\sqrt{s}$.
\begin{itemize}
\item[a,d)] parton densities given by set a: eq. (\ref{g=0}).
\item[b,e)] parton densities given by set b: eq. (\ref{g=max}).
\item[c,f)] parton densities given by set c: eq. (\ref{g=mid}).
\end{itemize}
The $Q^2$ range is restricted to $5 \mbox{ GeV}^2 < Q^2 < s$
and the cut-off values in (\ref{sijpij}) are $y_{c}=0.04$ and
$M_0= 2$ GeV.
\item[{\bf Fig. 4}]
$x$-dependence of the $O(\alpha_s)$ 3-jet processes at $\sqrt{s}=20$ GeV for
each subprocess and the sum (solid line) of all contributing subprocesses.
\begin{itemize}
\item[a)] parton densities given by set a: eq. (\ref{g=0}).
\item[b)] parton densities given by set b: eq. (\ref{g=max}).
\item[c)] parton densities given by set c: eq. (\ref{g=mid}).
\end{itemize}
The $Q^2$ range is restricted to $5 \mbox{ GeV}^2 < Q^2 < 7 \mbox{ GeV}^2$
and the cut-off values in (\ref{sijpij}) are $y_{c}=0.04$ and
$M_0= 2$ GeV.
\item[{\bf Fig. 5}]
same as Fig. 4 for $9 \mbox{ GeV}^2 < Q^2 < 11 \mbox{ GeV}^2$.
\item[{\bf Fig. 6}]
$W$-dependence of the $O(\alpha_s)$ 3-jet processes at $\sqrt{s}=20$ GeV for
each subprocess and the sum (solid line) of all contributing subprocesses.
\begin{itemize}
\item[a)] parton densities given by set a: eq. (\ref{g=0}).
\item[b)] parton densities given by set b: eq. (\ref{g=max}).
\item[c)] parton densities given by set c: eq. (\ref{g=mid}).
\end{itemize}
The $Q^2$ range is restricted to $5 \mbox{ GeV}^2 < Q^2 < 7 \mbox{ GeV}^2$
and the cut-off values in (\ref{sijpij}) are $y_{c}=0.04$ and
$M_0= 2$ GeV.
\item[{\bf Fig. 7}]
same as Fig. 6 for $9 \mbox{ GeV}^2 < Q^2 < 11 \mbox{ GeV}^2$.
\item[{\bf Fig. 8}]
same as fig. 4 for the
 $O(\alpha_s^2)$ 4-jet processes.
\item[{\bf Fig. 9}]
same as fig. 6 for the
 $O(\alpha_s^2)$ 4-jet processes.
\end{itemize}

\clearpage


\clearpage

\end{document}